# Assigning Creative Commons Licenses to Research Metadata: Issues and Cases

Marta POBLET [a,1], Amir ARYANI[b], Paolo MANGHI[c], Kathryn UNSWORTH[b], Jingbo WANG[d], Brigitte HAUSSTEIN[e], Sunje DALLMEIER-TIESSEN[f], Claus-Peter KLAS[e], Pompeu CASANOVAS [g,h] and Victor RODRIGUEZ-DONCEL[i]

[a] *RMIT University*
[b] *Australian National Data Service (ANDS)*
[c] *ISTI, Italian Research Council*
[d] *Australian National University*
[e] *GESIS – Leibniz Institute for the Social Sciences*
[f] *CERN*
[g] *IDT, Autonomous University of Barcelona*
[h] *Deakin University*
[i] *Universidad Politécnica de Madrid*

**Abstract.** This paper discusses the problem of lack of clear licensing and transparency of usage terms and conditions for research metadata. Making research data connected, discoverable and reusable are the key enablers of the new data revolution in research. We discuss how the lack of transparency hinders discovery of research data and make it disconnected from the publication and other trusted research outcomes. In addition, we discuss the application of Creative Commons licenses for research metadata, and provide some examples of the applicability of this approach to internationally known data infrastructures.

## Introduction

The emerging paradigm of open science relies on increased discovery, access, and sharing of trusted and open research data. New data infrastructures, policies, principles, and standards already provide the bases for data-driven research. For example, the FAIR Guiding Principles for scientific data management and stewardship [23] describe the four principles—findability, accessibility, interoperability, and reusability—that should inform how research data are produced, curated, shared, and stored. The same principles are applicable to metadata records, since they describe datasets and related research information (e.g. publications, grants, and contributors) that are essential for data discovery and management. Research metadata are an essential component of the open science ecosystem and, as stated in [17], "for a molecule of research metadata to move effectively between systems, the contextual information around it - the things that are linked to, must also be openly and persistently available".

---

[1] Corresponding Author: marta.pobletbalcell@rmit.edu.au

Yet, finding relevant, trusted, and reusable datasets remains a challenge for many researchers and their organisations. New discovery services address this issue by drawing on open public information, but the lack of transparency about legal licenses and terms of use for metadata records compromises their reuse. If licenses and terms of use are absent or ambiguous, discovery services lack basic information on how metadata records can be used, to what extent they can be transformed or augmented, or whether they can be utilised as part of commercial applications. Ultimately, legal uncertainty hinders investment and innovation in this domain.

The rest of this paper is organised as follows: Section 1 presents the most widely adopted research metadata protocols and practices; Section 2 provides some global figures about the types of licenses used for research metadata; Section 3 identifies the main stakeholders; Section 4 reviews the most common choices for metadata licenses and discusses both advantages and disadvantages of such choices; Section 5 offers six compact case studies from different research data services. Finally, the conclusion raises some questions to guide future work.

## 1. Research metadata protocols and practices

A number of instruments covering the management of research metadata are currently available. For example, the Open Archives Initiative (OAI) developed the Protocol for Metadata Harvesting OAI-PMH to facilitate interoperability between repositories and metadata service providers [14]. OAI-PMH enables harvesting the metadata of open access repositories such as PubMed, Arxiv, HAL, the Wikipedia [5], or the World Bank's Open Knowledge Repository (OKR).

The Dublin Core Metadata Initiative (DCMI) promotes interoperability and reusability in metadata design and best practices by developing semantic standards and recommendations, model-based specifications, and syntax guidelines, such as the Singapore Framework for Dublin Core Application Profiles or the DCMI Abstract Model.[2]

The RIOXX Metadata Guidelines,[3] implemented by more than 50 institutional repositories in the UK [22], have adopted NISO's "Recommended Practice on Metadata Indicators for Accessibility and Licensing of E-Content"[4] to add a tag (<license_ref>) with a reference to a URI carrying the license terms [13]. The main goal is to provide a mechanism of compliance with the RCUK policy on open access.

While the adoption of these instruments paves the way for technical standardisation, the discussion subsists with regard to the licensing options available and the implications of such choices. The issues arise out of the complexity of contractual obligations that the different types of licenses create, the extent of copyright laws in different jurisdictions, or the difficulties of attribution when metadata are combined or remixed. Since a distinctive feature of high-quality metadata is that "it is created once and then reused as needed" [3], transparency and predictability are essential. The available options come with different requirements, conditions, and scope.

---

[2] http://dublincore.org/specifications/
[3] http://rioxx.net/
[4] http://www.niso.org/apps/group_public/project/details.php?project_id=118

## 2. Current use of Creative Commons (CC) licenses in research metadata

Although many scientific data repositories live behind firewalls in proprietary environments, the Web houses thousands of scientific data repositories whose study is now possible. Marcial *et al*. [12] manually chose 100 diverse scientific data repositories and analysed 50 of their characteristics. While copyright issues were out of their scope, two of the observations referred to the input and output metadata's rights–distinguishing between the terms a contributor has to accept before uploading a new record and the license under which the entire metadata collection is offered. The text excerpts included in the 100 data repositories referring to these matters showed a huge variety of custom-made licenses and only two mentions to CC licenses were reported. The earliest data for this study were collected in 2007 and there is some evidence that the use of standardized licenses has dramatically increased since then. Yet, an updated study is still needed.

The Registry of Research Data Repositories by re3data.org (a service of DataCite) makes its data available for research under an API [15]. The Registry, now "the largest and most comprehensive registry of data repositories available on the web" [15] publishes an overview of existing international repositories for research data from all academic disciplines and as of September 2016, listed 1692 data repositories. An analysis of these repositories reveals that 269 (16%) of repositories made an explicit mention to CC licenses with a valid URI, while only 17 to Open Data Commons or 9 to GNU licenses. While these data require some caution (for example, the World Bank's Open Knowledge repository applies CC-BY 3.0 in most cases but it does not provide the corresponding URI) they offer a good snapshot of the current adoption of CC licenses in the research metadata ecosystem.

## 3. Main stakeholders

The following stakeholders can benefit from assigning Creative Commons (CC) licenses to the research public metadata:

- Research Management Software Vendors: Assigning CC licenses to research public metadata will encourage software vendors to incorporate this data into their systems, leading to better automation in data entry and discovery capabilities of research management systems.
- Research Institutions: Better research management systems can reduce the cost of data entry for universities, and enable discovery of research collaboration opportunities. In addition, universities will be able to demonstrate their collaboration network on the public domain using derivative analytics from CC licensed research metadata.
- Research Infrastructures (including data repositories): research metadata are the key enablers in creating interoperability between research infrastructures; particularly for research data repositories, public metadata enables connecting

- datasets across multiple systems and enables better discovery and reuse of the research output.
- Researchers: At present, finding related and relevant research, research data and other scholarly works is not a trivial task for most researchers. Better discovery tools augmented with public metadata would enable researchers to find related research and research collaborators, hence finding new research opportunities.
- Publishers: A clear indication on the applicable CC licenses would help to eliminate the uncertainties about possible consequences of reusing/republishing metadata.
- Funders: The collective effort by universities, publishers, infrastructure providers and software vendors can enable funders to have a better understanding of the impact of their funding; moreover, better research collaboration discovery can improve the return on investment.

**4. Applying Creative Commons Licenses to Research Metadata**

We address the issue of assigning clear licenses and terms of use for public research information by reviewing two of the most used assigning Creative Commons (CC) licenses for public metadata records: CC0 and CC-BY. Creative Commons discourages the use of its NonCommercial (NC) or NoDerivatives (ND) licenses on databases intended for scholarly or scientific use, and they are not open licenses according to the definition of 'open" by the Open Knowledge Foundation.[5] It is important to note that CC0 and CC-BY are not the only open licenses available, as Open Data Commons, to refer to another popular option, offers three legal tools –the Public Domain Dedication and License (PDDL), the Attribution License (ODL-By) and the Open Database License (ODBL)—which cover the European sui generis database right (although now this is also the case of CC 4.0 licenses). The choices will depend on the objects to be licensed (creative contents, data, databases, etc.), the clauses and terminology that come with each choice, the derived contractual obligations, and the mechanisms of enforcement available to the licensor.

The most accessible form of CC instrument is CC0 — "No Rights Reserved" (also known as Public Domain Dedication).[6] This is the choice of research data services such as Dryad or Figshare for their generated metadata. Increasingly, a number of cultural institutions such as the Tate Gallery, the Museum of Modern Art (MoMA), the Walters Art Museum, or the Thyssen Foundation are also releasing their metadata with the CC0 document.

Nevertheless, there are some doubts about the force of the CC0 waiver in some jurisdictions (e.g. under Australian law), especially with regard to moral rights. As AusGOAL alerts, "the disclaimer that accompanies CC0, at present, may be ineffective in protecting the user from liability for claims of negligence." The main issue with assigning a CC0 document to research metadata is the responsibility to collect the original records with the CC0 waiver. According to the Creative Commons definition (CC0 2016), "You should only apply CC0 to your own work, unless you have the necessary rights to apply CC0 to another person's work."[7] Hence, unless adequate provisions are taken, metadata aggregators or repositories would not be able to assign the CC0 license to records created by other sources. This is why, for instance, Europeana releases all its metadata with the CC0 document and requires its data providers to waive all IP rights to the metadata provided. Likewise the Digital Public Library of America (DPLA) requires all data and metadata donors to attach a CC0 document to any donation [7].

---

[5] http://opendefinition.org
[6] https://creativecommons.org/about/cc0

Another popular CC tool for open access works is the CC-BY license that enables third parties to distribute the work with attribution to the original author. The problem of assigning CC-BY licenses to aggregated metadata is that the sources of metadata records are not always clear. Who owns metadata records? The researcher who described the work? The research institution who owns the IP? Moreover, the CC-BY license requires to "indicate if changes were made" which adds to the complexity of enriching metadata by aggregators (CC-BY 2016). Given these options, assuming Copyright in metadata seems to be the safest approach. As AusGOAL advises, "recent developments in Australia have led to the situation where it is unclear which data is subject to copyright. In this situation, Australian researchers have to take a pragmatic approach and it would seem desirable to assume copyright as subsisting in all data created in the course of research, and ensure that it is licensed accordingly. No harm can come from this approach." [4]. ANDS adds to this, "It will still serve as a useful way to make known how you would like to be attributed, in addition to applying a limitation of liability and warranty clause to the data" [1]. In cases where it is clear that copyright does not subsist in the aggregated metadata, applying a CC Public Domain mark would suffice, provided the rights to do so have been established, including consideration that copyright for the material may subsist in other jurisdictions.

## 5. Case studies

In addition to the global numbers provided in Section 2, a small set of brief case studies on the application of Creative Commons licenses for metadata was made including ANDS, CERN, da|ra, NCI, OpenAIRE, and Research Graph. Even if limited, this information can shed some light into the applicability of CC licenses to research data infrastructures.

- **ANDS:** The Australian National Data Service (ANDS)[8] manages Research Data Australia (RDA), a national research data registry. RDA receives contributions from more than 100 Australian research institutions, data infrastructures, and research organisations (RDA 2016). ANDS collects and publishes all metadata under an agreement with contributors by which their records will be openly available on the web (ANDS Agreement 2010); however, there is no license attached to these metadata records, as often contributors do not assign licenses to their records.
- **CERN**: The European Organization for Nuclear Research[9] has different platforms and services related to scholarly information. CERN offers numerous platforms and services related to scholarly information. For the purpose of this case study, two of them should be highlighted here: (i) INSPIREHEP, the main information platform in high-energy physics, aggregates scholarly information from all relevant community resources. On top of it, the service provides 'author pages' (with ORCID integration) compiling information about researchers from the scholarly records available on INSPIRE. The metadata on this platform are shared with a CC0 waiver, with the expectation that third parties will use the available information to compile new services, such as citation statistics; (ii) the Open Data Portal publishes data and research materials accompanying datasets: documentation, software, trigger files, and tutorials to enable reuse by any interested audience. Objects are shared with Open Science licences, data and metadata with the CC0 mark, and software with the GNU General Public License (GPL).

---

[7] https://creativecommons.org/share-your-work/public-domain/cc0/
[8] http://www.ands.org.au/
[9] https://home.cern/

- **da|ra: da|ra is a** registration agency for social science and economics data in Germany.[10] It is run by the GESIS Leibniz Institute for the Social Sciences and ZBW Leibniz Information Center for Economics, in cooperation with DataCite (the international consortium promoting research data as independent citable scientific objects). This infrastructure lays the foundation for long-term, persistent identification, storage, localization and reliable citation of research data via allocation of DOI names. Each DOI name is linked to a set of metadata and presents the properties of resources, their structure and contextual relations. The da|ra Metadata Schema [10] provides a number of mandatory elements – core properties –that have to be submitted by the publication agent at the time of data registration. Publication agents may also choose other optional properties to identify their data. Although da|ra complies with the official DataCite Metadata Schema, it has broadened the DataCite metadata by adding some specific properties related to the social sciences and economics. da|ra reserves the right to share metadata with information indexes and other entities. da|ra supports the open metadata principles and all metadata are available under CC0 1.0. to encourage all metadata providers (data centers, data repositories, libraries, etc.) to make their metadata available under the same terms. Since 2016 da|ra has been offering access to the metadata of the registered research data using the Open Archives Initiative Protocol for Metadata Harvesting [14]. The da|ra OAI-PMH Data Provider is able to disseminate records in the following formats: DDI-Lifecycle 3.1 and OAI DC.

- **NCI**: The National Computational Infrastructure (NCI)[11] at the Australian National University (ANU) has evolved to become Australia's peak computing centre for national computational and Data-intensive Earth system science. More recently NCI collocated 10 Petabytes of 34 major national and international environmental, climate, earth system, geophysics and astronomy data collections to create the National Environmental Research Interoperability Data Platform (NERDIP). Data Collection management has become an essential activity at NCI. NCI's partners (CSIRO, Bureau of Meteorology, Australian National University, and Geoscience Australia), supported by the Australian Government and Research Data Storage Infrastructure (RDSI) and Research Data Services (RDS), have established a national data resource that is co-located with high-performance computing. Most of the data are quality assured for being 'published' and made accessible as services under Creative Commons Attribution (CC BY) 4.0 as they are sourced from government agencies [10]. The license files are published jointly with data through NCI's OpenDAP server (http://dap.nci.org.au). The metadata associated with data collection are available under CCBY4.0. They are publicly available for users to query the metadata catalogue entries. Our collection level metadata has also been harvested by national and international aggregators such as RDA and International Directory Network of Committee on Earth Observation Satellites (CEOS).

---

[10] http://www.da-ra.de/
[11] http://nci.org.au

- **OpenAIRE**: The OpenAIRE[12] infrastructure is the point of reference for Open Access and Open Science in Europe (and beyond) [9]. Its mission is twofold: enabling the Open Science cultural shift of the current scientific communication infrastructure by linking, engaging, and aligning people, ideas, resources, and services at the global level; monitoring of Open Access trends and measuring research impact in terms of publications and datasets to serve research communities and funders. To this aim, OpenAIRE offers services [21] that collect, harmonize, de-duplicate, and enrich by inference (text mining) or end-user feedback, metadata relative to publications, datasets, organizations, persons, projects and several funders from all over the world. Starting 2017, in order to join the infrastructure, data sources will sign a Terms of Agreement where they will grant to the OpenAIRE services the right of collecting and reusing metadata records under CC-0. This is expected to impact on the general trend of institutional and thematic publications repositories, which today are not exposing any license metadata together with their APIs. From a recent analysis, out of a sample of around 2500 publication repository services in OpenDOAR 2 (supporting the OAI-PMH protocol standard), only 9 expose metadata license information: 3 with CC-0, 2 with CC-BY, and 4 which require a permission for commercial use, 3 with CC-0 and 1 with CC-BY. The graph is exported via standard protocols (e.g. HTTP-REST search, LOD, OAI-PMH) and formats, and the metadata records are available under CC-BY or CC-0, with no restriction of embargo or re-use.
- **Research Graph**: This is an example of value added to data infrastructures by third-party services. Research Graph[13] is a collaborative project by a number of international partners that links research information (datasets, grants, publications and researchers) across multiple platforms. This initiative uses the research metadata to construct a graph of scholarly works, and this graph connects data and publications with multiple degrees of separation. The outcome enables a researcher to search the graph for a particular publication or research project and discovers a collaboration network of researchers who are connected to this work (or topic). The main consideration for such a service is the ability to read, connect and transform metadata, and without clear licensing or terms of use this platform would not be able to include a data infrastructure in the graph. In addition, given the mixture of licences provided by different contributors the constructed graph is not publically available as a single dataset. Instead, it had to be split into separate clusters where the license for each cluster can be defined from the individual sources.

---

[12] https://www.openaire.eu/
[13] http://researchgraph.org/

## 6. Conclusions and future work

In this paper we have raised the need for a transparent regime of metadata licensing and have briefly reviewed the state-of-the art application of open licenses to research metadata. We have offered some global figures on the use of CC licenses in scientific research metadata, explored the differences between CC0 and CC-BY, and the approach taken by six data registries and/or repositories.

The use of standardized licenses fosters reusability and science in general, and choosing CC or alternative well-known licenses favours the automatic discovery of usable datasets (e.g. *search by license*), which can be accomplished by means of the re3data.org API. An additional advantage of CC licenses is their availability in a machine-readable form, namely, the license document contains a RDFa which enable further intelligent processing of the license content (e.g. search by specific conditions).

The examples reviewed did not include research funding bodies, an integral part of the meta-research ecosystem [16] and the integration of researcher identifiers (such as ORCID); hence, we believe that these are areas for future investigation. Likewise, it could be further investigated if research metadata, as compared to other types of metadata, have enough specificity to require a dedicated set of licenses.

Data licensing has attracted the attention of many researchers in the fields of Semantic Web, computational linguistics and deontic logic. Datasets of RDF licenses do exist already [20]. The challenge of developing automated frameworks able to generate licensing data terms from heterogeneous distributed sources has also been addressed [9]. NLP techniques to extract rights and conditions granted by licenses and return them into RDF have been applied [6]. We believe that these technical solutions offer a number of advantages and deserve to be monitored, reused, and tested in real scenarios.

However, the context-dependent problems and ambiguities highlighted in this paper still survive, such as the discussion of how to share the metadata generated in scientific research, whether it should accrue the public domain or, rather, whether scientists and research organisations should retain a legal expression of attribution, including the implications of such choices. For example, if assigning CC licenses to remixed metadata requires modifying the records at the source (or get consent from all creators involved), the process can be difficult to escalate.

This is a domain that requires some *previous* positioning before making decisions about the appropriate heuristics for end-users interfaces. The underlying philosophy and assumptions about building semi-automated ecosystems for data science face practical legal, commercial, and political issues that require attention.